\documentclass[manuscript]{aastex}	
\usepackage{epsfig}			

\usepackage{graphicx,color}		
\usepackage{amssymb}
\usepackage{color}			
\usepackage{url}			
\usepackage{amsmath}			
\usepackage{rotating}			
\usepackage{float}			
\usepackage{textcomp}			
\usepackage{psfig}
\usepackage{dcolumn}
\usepackage{times}
\usepackage{tabularx}
\usepackage[bottom]{footmisc}
\usepackage{tablefootnote}
\usepackage[colorlinks=true,citecolor=blue]{hyperref}
\usepackage[english]{babel}
\usepackage{lineno}
\usepackage{ragged2e}
\justifying

\begin{document}
\title{Homogenising SoHO/EIT and SDO/AIA 171\AA~ Images: A Deep Learning Approach}

\author{Subhamoy Chatterjee$^{1}$, 
Andr\'{e}s Mu\~{n}oz-Jaramillo$^{1}$, 
Maher Dayeh$^{2,3}$, 
Hazel M. Bain$^{4,5}$,  
Kimberly Moreland$^{2,3,4,5}$}

\affil{$^{1}$ Southwest Research Institute, Boulder, CO, USA.\\
$^{2}$ Southwest Research Institute, San Antonio, TX, USA.\\
$^{3}$ The University of Texas at San Antonio, San Antonio, TX, USA.\\
$^{4}$ Cooperative Institute for Research in Environmental Sciences, University of Boulder, CO, USA.\\
$^{5}$ Space Weather Prediction Center, NOAA, Boulder, CO, USA.\\
e-mail: {\color{blue}{subhamoy.chatterjee@swri.org}}
}

\begin{abstract}
Extreme Ultraviolet images of the Sun are becoming an integral part of space weather prediction tasks.  However, having different surveys requires the development of instrument-specific prediction algorithms.  As an alternative, it is possible to combine multiple surveys to create a homogeneous dataset. In this study, we utilize the temporal overlap of SoHO/EIT and SDO/AIA 171~\AA ~surveys to train an ensemble of deep learning models for creating a single homogeneous survey of EUV images for 2 solar cycles.   Prior applications of deep learning have focused on validating the homogeneity of the output while overlooking the systematic estimation of uncertainty. We use an approach called `Approximate Bayesian Ensembling' to generate an ensemble of models whose uncertainty mimics that of a fully Bayesian neural network at a fraction of the cost.  We find that ensemble uncertainty goes down as the training set size increases. Additionally, we show that the model ensemble adds immense value to the prediction by showing higher uncertainty in test data that are not well represented in the training data. 
\end{abstract}


\section{INTRODUCTION}
    Solar observations span across several past decades being recorded by multiple ground-based and space-based observatories. Due to changes in instrumentation, the datasets differ in resolution (spatial and temporal), field-of-view, dynamic range, and noise characteristics. While dealing with long-term solar data it requires building custom algorithms to detect solar features/events for each instrument. Cross-calibrating those surveys and creating a single homogeneous dataset for the entire span of observation enables the scientific community to make long-term studies and discover underlying patterns without spending effort to deal with instrument differences.\\
    
    There have been several efforts to homogenize solar images from different surveys using traditional approaches such as oversampling of pixels using interpolation, intensity rescaling etc. Recent studies find a significant improvement over those baseline approaches in homogenization tasks such as super-resolving solar magnetograms \citep{2019arXiv191101490J}  utilizing state-of-the-art machine learning (ML) approaches such as Convolutional Neural Networks. \\
    
    Deep-learning approaches applied in scientific domains suffer from the shortcoming of limited availability of data. Limited data creates barriers for ML models to learn the diversity of natural phenomena. Thus, in such situations trusting a point estimate by an ML model on unseen data can result in underestimation and overestimation. Creating model-ensembles can help mitigate this problem by generating prediction uncertainty. There are different approaches to creating deep learning model-ensembles that range from multiple selections of data to random disconnections in neural networks \citep{2015arXiv150602142G} and approximating Bayesian Neural Networks by adding several anchors that drive the regularization term in the model loss \citep{2018arXiv181005546P}. Studies find the application of usage model-ensembles to be useful in overcoming the demerits of data scarcity in scientific domains such as heliophysics \citep{2022NatAs...6..796C}\\
    
    In this study, we use convolutional neural networks to homogenize and estimate the uncertainty of, full-disc solar extreme ultraviolet image in 171\AA~from Extreme-ultraviolet Imaging Telescope (EIT;\citet{1995SoPh..162..291D}) onboard Solar and Heliospheric Observatory (SoHO) and Atmospheric Imaging Assembly (AIA;\citet{2012SoPh..275...17L}) onboard Solar Dynamics Observatory (SDO) using their overlapping period.

\section{OBSERVATIONAL DATA}
\begin{figure}[htbp!]
\hspace{-0.05\textwidth}
\includegraphics[width=1.1\linewidth]{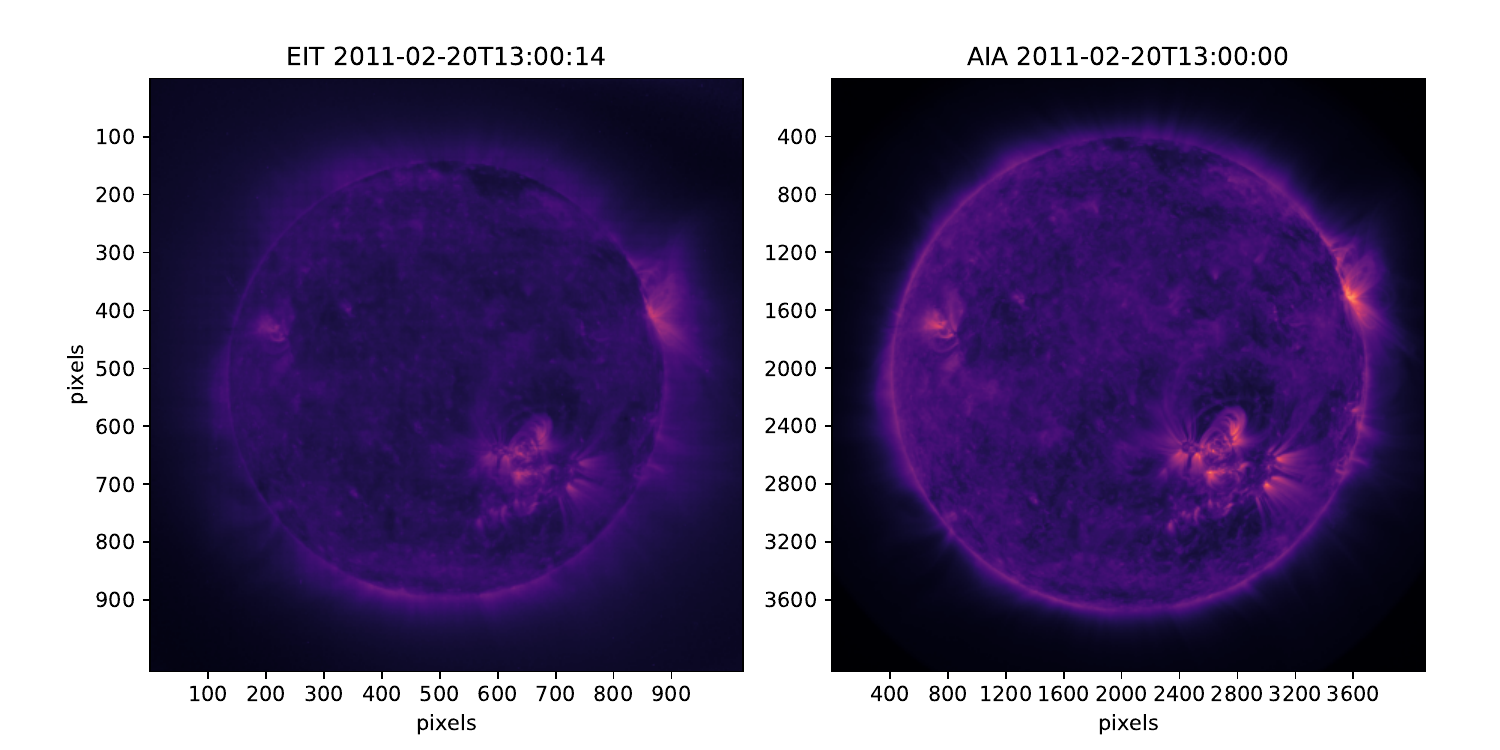}
\caption{Level-1 representative images from SoHO/EIT and SDO/AIA from the same date. \textit{Left panel} shows an EIT 171\AA~image at the native resolution and field-of-view. \textit{right panel} shows an AIA 171\AA~image at native resolution and field-of-view (FoV). A comparison of the images clearly depicts the higher FoV of EIT.}
\label{fig:lev1}
\end{figure}

We use level-1 SoHO/EIT and SDO/AIA 171\AA~ full-disc images (Figure~\ref{fig:lev1}) with a cadence of 1 day and over the period 2010-2020 for our study. We make sure that the EIT-AIA pairs are observed at the same time.

\section{DATA ALIGNMENT}

\subsection{Re-projection}
\begin{figure}[htbp!]
\hspace{-0.05\textwidth}
\includegraphics[width=1.1\linewidth]{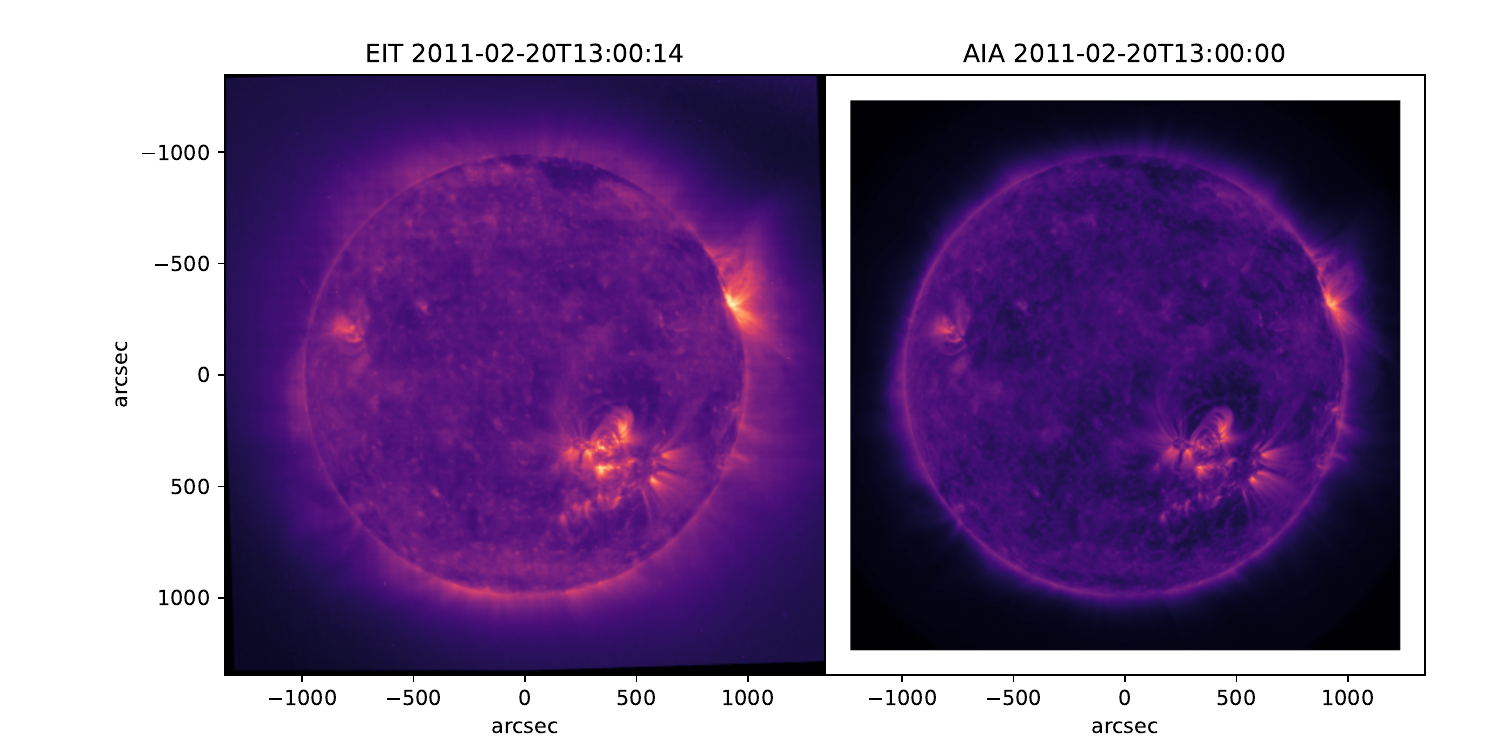}
\caption{Outcome of re-projection. \textit{Left panel} shows a EIT 171~\AA ~image; \textit{Right panel} shows a re-projected AIA image to match the FoV and point-of-view seen by EIT detector.}
\label{fig:reproj}
\end{figure}

The EIT-Sun-AIA angle, focal length and detector differences necessitate re-projecting EIT, AIA images to a common point-of-view and field-of-view. We use the `reproject' module of Python to map AIA images to EIT detector with 4 times the original resolution of EIT (Figure~\ref{fig:reproj}). The re-projection of the corona is non-trivial. So, we make an assumption that coronal structures lie on a sphere centered on Earth, with a radius of Sun-Earth distance.

\subsection{Template matching}
\begin{figure}[htbp!]
\hspace{-0.05\textwidth}
\includegraphics[width=1.1\linewidth]{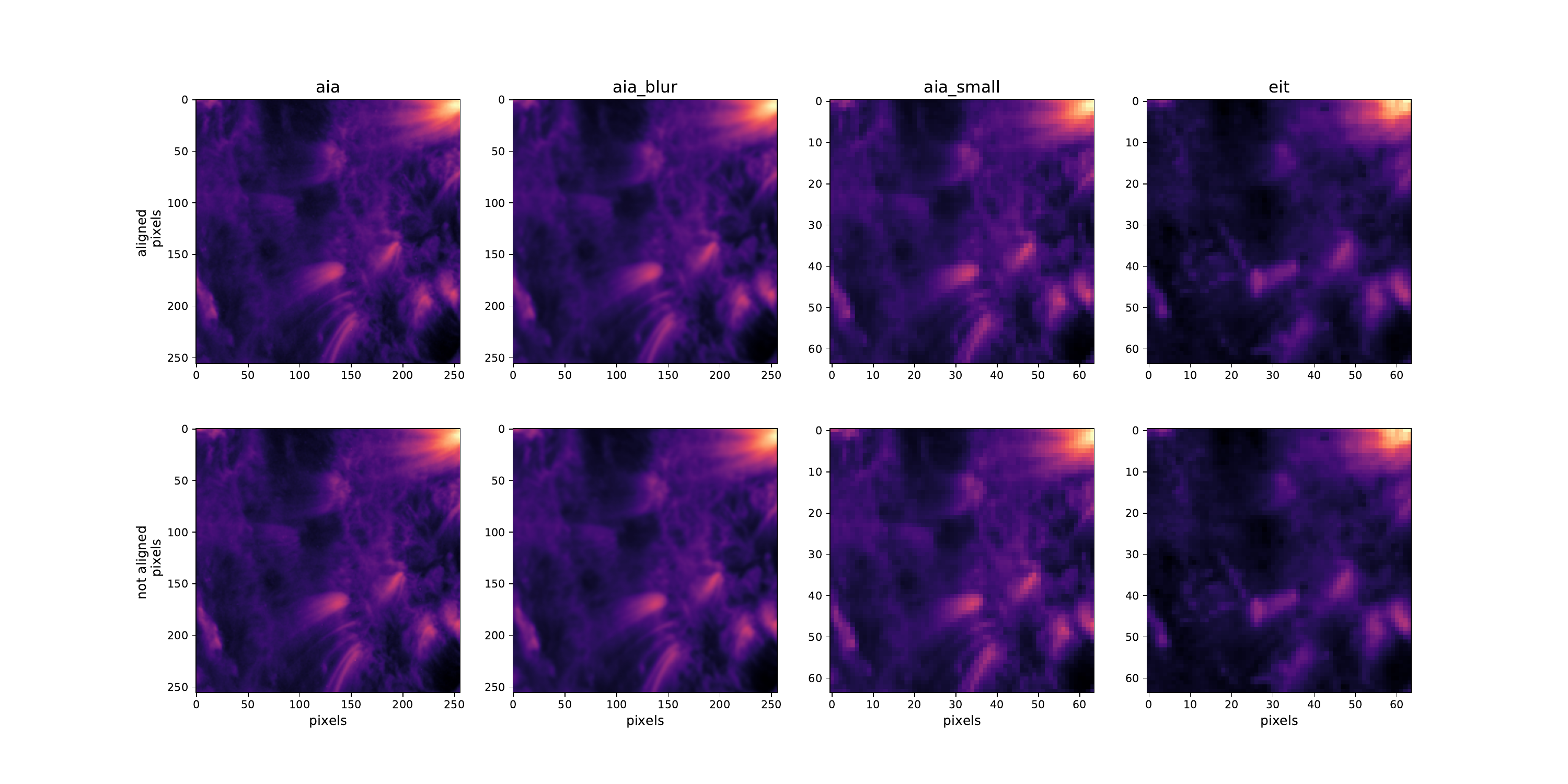}
\caption{Benefit of patch alignment while creating input-target pair for EIT-to-AIA transformation. \textit{Top row} shows the effect of patch alignment while the \textit{bottom row} shows unaligned patches. While comparing the unaligned AIA patches with the aligned ones, the vertical and horizontal shift correction becomes apparent by looking at the bright region on top-right. Different columns for the AIA patch represent steps toward undersampling for visual comparison with EIT at native resolution.}
\label{fig:align}
\end{figure}

Re-projection is insufficient when we want to perform one-to-one mapping of EIT to AIA patches.
Several other factors such as resolution, and optical aberration result in pixel shifts and may affect our homogenization objective. To make an accurate alignment we perform template matching \citep{10.5555/1643435} using the following steps-
\begin{enumerate}
    \item Divide the EIT full disc images into contiguous non-overlapping $64\times64$ patches.
    \item Blow up each of those patches by 4 times and map that location to the re-projected AIA.
    \item Correlate that blown-up EIT patch within the corresponding AIA location having a margin of (1/8)$^{th}$ patch size on all 4 sides.
    \item Find the 256$\times$256 region from the AIA window providing the best correlation.
    \item Follow the above steps after generating another set of non-overlapping patches bounded by the center pixels of the original patches.
\end{enumerate}
The steps above produce 365 patch-pairs [64$\times$64 EIT, 256$\times$256 AIA] per full-disc EIT-AIA image-pair (Figure~\ref{fig:align}).

\subsection{Data preparation: training, validation and test set}
We sample decade-long data (05/2010- 12/2020) within the training, validation, and test set by adding the first 7 months of each year to the training set, August and September to the validation set, and, the rest to the test set. We also put the entire 2015 data aside in the test set and tune the model hyperparameters based on the validation set performance. 

\section{DL-based IMAGE TRANSFORMATION}

\subsection{Deep learning model}
We prepare the Deep learning (DL) model to transform 64$\times$64 EIT patches to 256$\times$256 AIA patches. 
The model is based on the super-resolution convolution neural net (CNN) presented in \citet{deudon2020highresnet}.
However, our network is simple as we don't have multiple views and avoid shift-net by pre-aligning the EIT, and AIA patches. The model consists of an encoder and a decoder structure. The encoder consists of 2D convolution layers and residual blocks. The decoder consists of 4x upsampling and 2D convolution layers. In each residual block, two convolution layers are used and the convolution layer outcome is added to the input layer to generate the output. We use a kernel size of $5\times5$, 32 feature layers and a parametric ReLU activation in all the convolution layers except for the final layer that has only one feature layer and a kernel size of $1\times1$. We use reflection padding the deal bring edge continuity in the adjacent super-resolved tiles.
\subsection{Loss function definition}
To minimize the difference between the target ($T$) and DL predicted ($P$) image we consider the following as elements of loss function-
\begin{enumerate}
\item Mean squared error: $MSE = \frac{1}{N}\sum_{i}{(T_i - P_i)^2}$ where $N$ stands for the number pixels in output patches.
\item Histogram \citep{DBLP:journals/corr/abs-1804-09398} difference: $H = \frac{1}{K}\sum_{i}{(HT_i - HP_i)^2}$ where $K$ stands for the number of bins in the histograms.
\item Negative of Structural Similarity Index Metric \citep{1284395}: $S = -\frac{(2\mu_T\mu_P+C_1)(2\sigma_{T,P}+C_2)}{(\mu_T^2 + \mu_P^2 + C_1)(\sigma_T^2 + \sigma_P^2 + C_2)}$ with $C_1 = (K_1L)^2, C_2 = (K_2L)^2$, $L$ being dynamic range and $K_1,K_2\ll1$. $\mu_X, \sigma_X, \sigma_{X,Y}$ respectively represent mean of $X$, standard deviation of $X$ and covariance of $X, Y$
\item Gradient difference: $G = \frac{1}{N}\sum_{i}{(\nabla_x T_i - \nabla_x P_i)^2} + \frac{1}{N}\sum_{i}{(\nabla_y T_i - \nabla_y P_i)^2}$
\end{enumerate}
Our objective is to perform the best performing weighted combination of $MSE$ and one of the other losses. 
To identify the weights we combine $MSE$ with each of the remaining components and calculate the best weight through a set of diagnostics as described in the following subsection.

\subsection{Approximate Bayesian Ensembling}
Having a small amount of training data it is important that we estimate the uncertainty in model inference. We use a state-of-the-art technique called Approximately Bayesian Ensembling (ABE) that draws a set of anchor weights 
from a prior distribution and factor them through an additional regularization term to the loss function defined by-
\begin{linenomath*}
    \begin{equation}
    Loss_{n} = \frac{1}{N}\lVert \mathbf{I} - \mathbf{\hat{I}}_n \rVert_{2}^2 + \frac{1}{N} \lVert \mathbf{\Gamma}^{1/2}.(\boldsymbol\theta_n - \boldsymbol\theta_{anchor,n}) \rVert_{2}^2
    \end{equation}
\end{linenomath*}
where $diag(\mathbf{\Gamma})_i = \frac{\sigma_{\epsilon}^{2}}{\sigma_{prior,i}^{2}}$ and $N$ is the number of datapoints\\
During the training phase, the weights ($\boldsymbol\theta_n$) are optimized but the anchors ($\boldsymbol\theta_{anchor,n}$) are kept unchanged. We randomly select a set of four anchors and optimize a model for each giving rise to the ensemble of models. Here the index $n$ runs over the ensemble members. We set $\sigma_{\epsilon}^{2}$ to be the mean of the AIA patch histogram. We use anchored regularization in the last layer of our model to select both priors and anchors from a Glorot uniform \citep{pmlr-v9-glorot10a} distribution with $\sigma_{prior,i}^{2} = \frac{6}{n_l + n_{l+1}}$ where $n_l$ and $n_{l+1}$ are the numbers of units in two consecutive layers. We rewrite $diag(\mathbf{\Gamma})_i$ as $\frac{k_1(n_l + n_{l+1})}{6}$. We use $k_1$ (with a default value of $\sigma_{\epsilon}^{2} \approx 0.5$) to change the regularization weight and another parameter $k_2$ (with a default value of 1) to scale the standard deviation of the anchor weight distribution. We examine the sensitivity of the results to the anchored regularization with different values of $k_1$ (later referred to as `LAMBDA') and $k_2$ (later referred to as `SD').

\section{Results}

\subsection{Model prediction, diagnostics and comparison with baseline approach}
We develop a set of diagnostics to compare the model inference between the baseline and outcomes for different loss function combinations. For this purpose, we collect all the validation patch pairs and model inference on those. We generate 2-dimensional histograms through logarithmic binning (uniform bins in log space) from the tuple of pixel intensity from model inference and target. We calculate the median and standard deviation for each inferred intensity bin. This generates a curve representing the expected AIA target and uncertainty for each inferred value. From those, we estimate the deviation from the expected rule i.e. `target = expected'. We name that metric as `\% relative residual'. We also add a metric called `\% relative variance' representing the standard deviation and median ratio.\\

\begin{figure}[htbp!]
\hspace{-0.18\textwidth}
\includegraphics[width=1.3\linewidth]{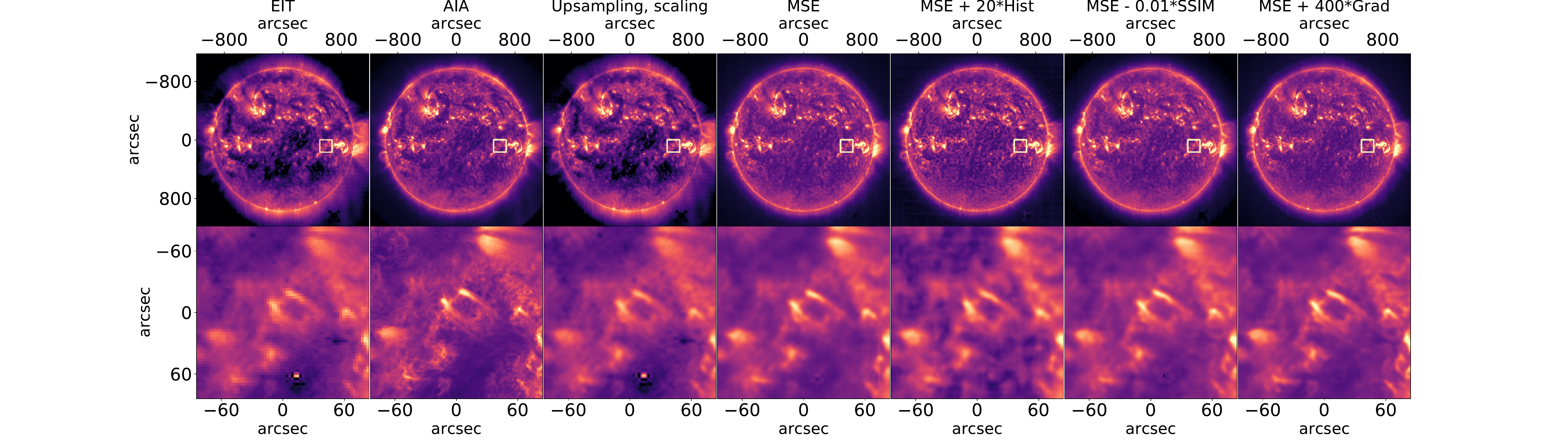}
\caption{Effect of loss function variations of inferred high-resolution image. The baseline outcome is depicted by the 3rd column while starting from the 4th column all the outcomes towards the right are from a DL model with loss function shown at the titles. The effect of loss function variation is highly apparent the model outcome. For example, MSE + 20*Hist does not show much micro-details as depicted by the MSE + 400*Grad. It can also be seen that the DL outcomes get rid of the apparent artifact (bright dot on patch bottom half) on EIT.}
\label{fig:infer}
\end{figure}

\begin{figure}[htbp!]
\hspace{-0.05\textwidth}
\includegraphics[width=1.1\linewidth]{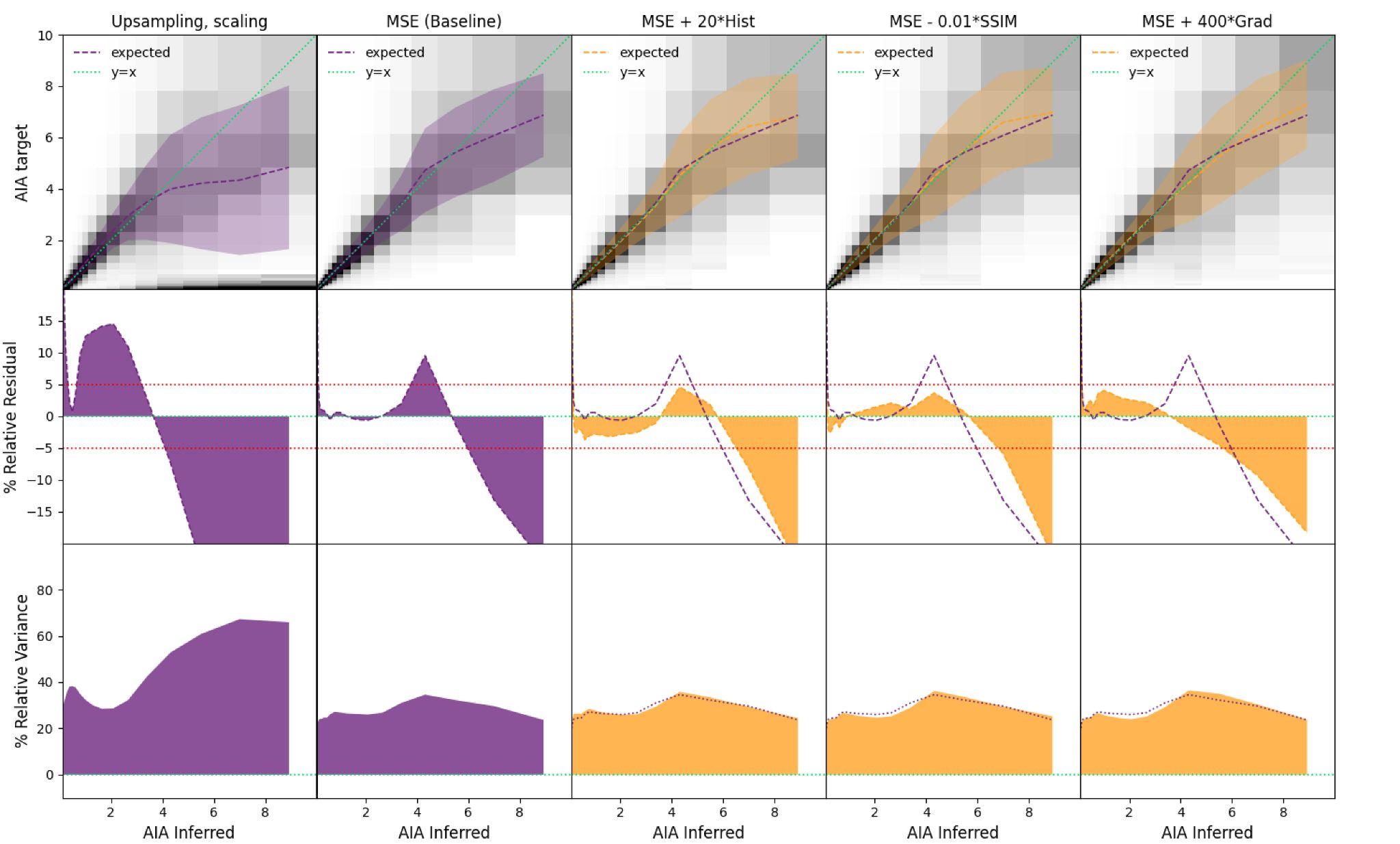}
\caption{Change in metrics defining the quality of inferred images with change in the loss function. The leftmost column shows the outcome of upsampling and scaling. 2nd to 5th column starting from left show the outcomes of the DL model acquired through minimizing a loss printed as the title of each column. The top row shows a 2D histogram between the target and inferred AIA intensities across all the test patches. The dashed lines represent the median line calculated through each input inferred intensity bin and the shaded region represents the inter-quartile range.  The green diagonal line represents `y=x'. The middle row depicts the relative residual for meaning percentage deviation of the median line from `y=x' line. The horizontal dotted lines mark +/- 5\% relative residual. The bottom row represents relative variance meaning the percentage uncertainty of outcome w.r.t. median. The outcome for MSE loss is overplotted as a baseline on all other loss function outcomes.}
\label{fig:metrics}
\end{figure}

From the visual inspection and metrics (Figures~\ref{fig:infer},\ref{fig:metrics}), we find that all the DL outcomes give rise to superior results as compared to those corresponding to simple Upsampling and scaling. As depicted in Figure~\ref{fig:infer}, the DL outcomes get rid of the bright point-like artifact present in the EIT patch. Among the DL outcomes, we use MSE loss as a baseline and examine the improvement caused by additional terms in the loss function. We find that unlike MSE additional loss terms drive the `\% relative residual' to be within $\pm 5\%$ for an inferred intensity range of [0,6] (Figure~\ref{fig:metrics}). We also observe that MSE + 400*Grad performs the best for higher intensities ($>6$). Also, by visual inspection, one can discern finer details in MSE+400*Grad more easily as compared to other DL outcomes. 

The results show that (although better than the baseline) the deep-learning model cannot reach the same level of detail as the target AIA 171~\AA images (Figure~\ref{fig:infer}). This could be because of the ill-posed nature of the problem with AIA having higher diffraction-limited resolution and better pixel pitch than EIT. It could also be the limited training data that is not able to constrain the outcome enough or that the small-scale information can't be uniquely recovered from the large scale. More experiments should be performed to understand the optimal number of loss function terms beyond which performance can't be improved given the size of the training data. Recent advancement in conditional GANs also inspires further experiments to bring back surface texture information but physical validation is needed for them to be fit for scientific applications. 

\subsection{Sensitivity of Uncertainty to Training set Size}

\begin{figure}[htbp!]
\hspace{-0.03\textwidth}
\includegraphics[width=1.05\linewidth]{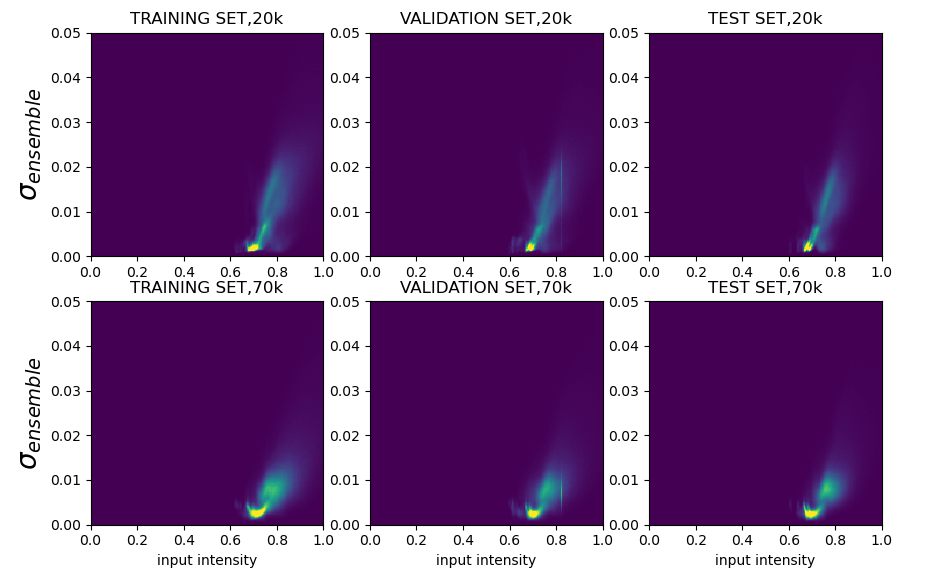}
\caption{Change in inference uncertainty with training set size. Each panel represents a 2D histogram between input pixel intensity and standard deviation of ensemble inference ($\sigma_{ensemble}$) across 4 super-resolved pixels. The top row represents the 2D histograms made from training, validation, and test patches when the models were trained with 20000 patch pairs. The bottom row represents the 2D histograms made from training, validation, and test patches when the models were trained with 70000 patch pairs. A clear reduction in ensemble SD can be observed especially for higher-intensity pixels when the training set gets bigger.}
\label{fig:uncert_1}
\end{figure}

We select four anchors and train a model with each of those without changing the training and validation set. 
Initially, we train the model with 20000 EIT-AIA patch pair and then increase the number of training patches to 70000. To record the change in performance, we generate 2-dimensional histograms between ensemble standard deviation vs. input EIT intensity over all the patch pixels for training, validation, and test sets. The ensemble standard deviation ($\sigma_{ensemble}$) is calculated over 64 pixels (4$\times$4 for 4 models) corresponding to each EIT pixel. From 2D histograms, we find a clear impact of training set size on the ensemble uncertainty. We observe that for each EIT intensity, the distribution of ensemble uncertainty when trained with 70000 patches shifts towards smaller values as compared to when trained with 20000 patches (Figure~\ref{fig:uncert_1}).\\

\subsection{Sensitivity of Uncertainty to ensembling parameters}

\begin{figure}[htbp!]
\vspace{-0.15\textwidth}
\hspace{-0.1\textwidth}
\includegraphics[width=1.2\linewidth]{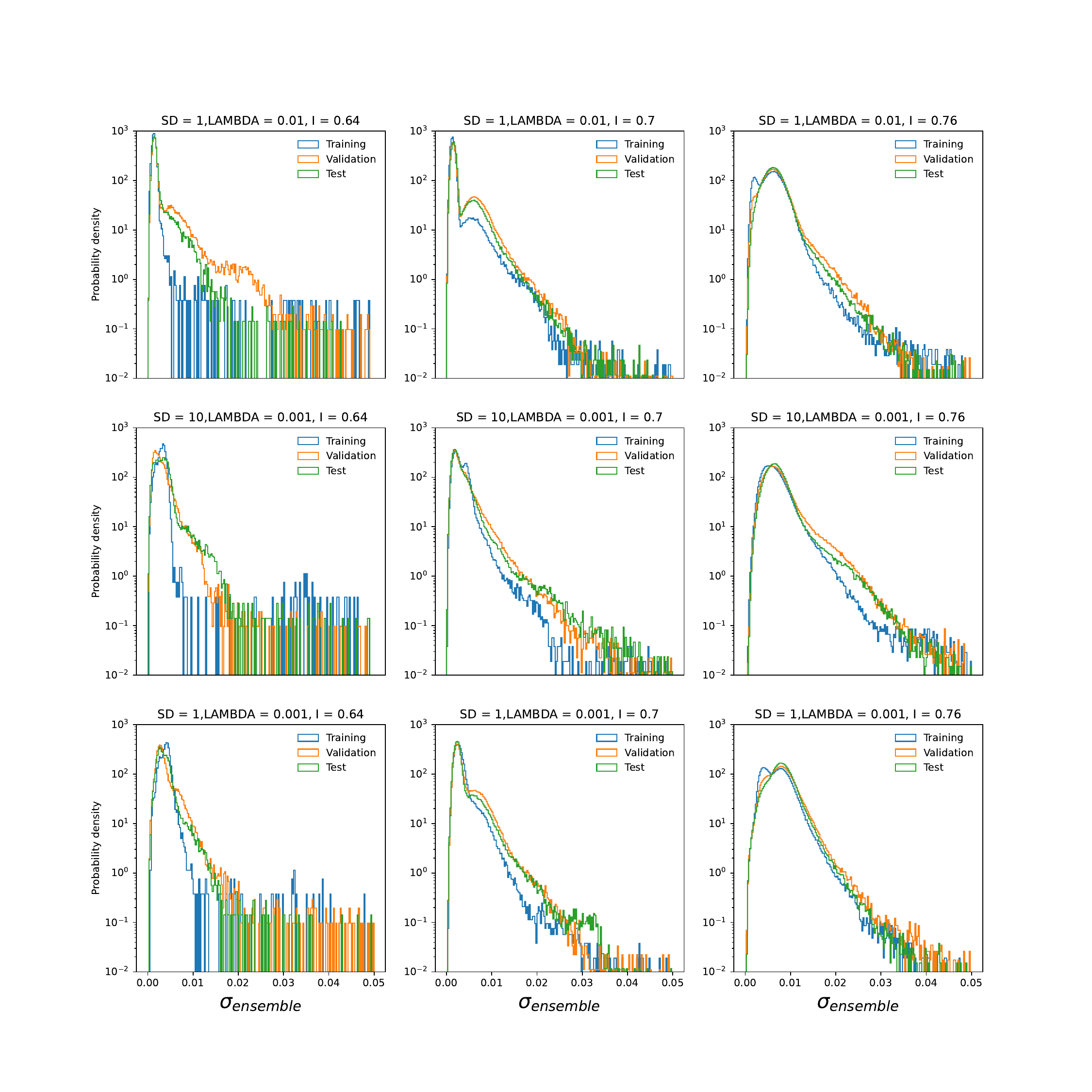}
\vspace{-0.15\textwidth}
\caption{Change in ensemble uncertainty ($\sigma_{ensemble}$) with the intensity of input pixels. Each panel shows uncertainty histograms on training, validation, and test sets for a particular input intensity and setting of the ensemble run. Each row represents the effect of a particular standard deviation scaling factor (SD) of the regularization term's anchor distribution and weightage (LAMBDA). Each column represents the outcome for a particular input intensity (I). The stronger difference in histogram tails can be observed for the lowest input intensity i.e. I = 0.64 with validation and test set outcomes being more extended towards higher uncertainty values.}
\label{fig:uncert_2}
\end{figure}

We also record changes in the distribution of ensemble standard deviation ($\sigma_{ensemble}$) for values of input intensity and parameters defining anchor distribution ($k_2$ or `SD') and weightage of the regularization term ($k_1$ or `LAMBDA') in the loss function over training, validation, and test set. It appears that for all the input intensities and anchor parameters, the $\sigma_{ensemble}$ distributions have a longer tail in the validation and test set as compared to the training set. The difference between validation/test and training set distribution diminishes for higher input intensities. However, the distributions get wider for higher input intensities. We don't find as pronounced an effect by changing the anchor parameters. However, a close inspection reveals that the distribution gets wider with a reduction in regularization weightage for I=0.64. Also, for I = 0.76 the second peak in the distribution gets less pronounced for lower weightage (Figure~\ref{fig:uncert_2}).

\begin{figure}[htbp!]
\hspace{-0.12\textwidth}
\includegraphics[width=1.2\linewidth]{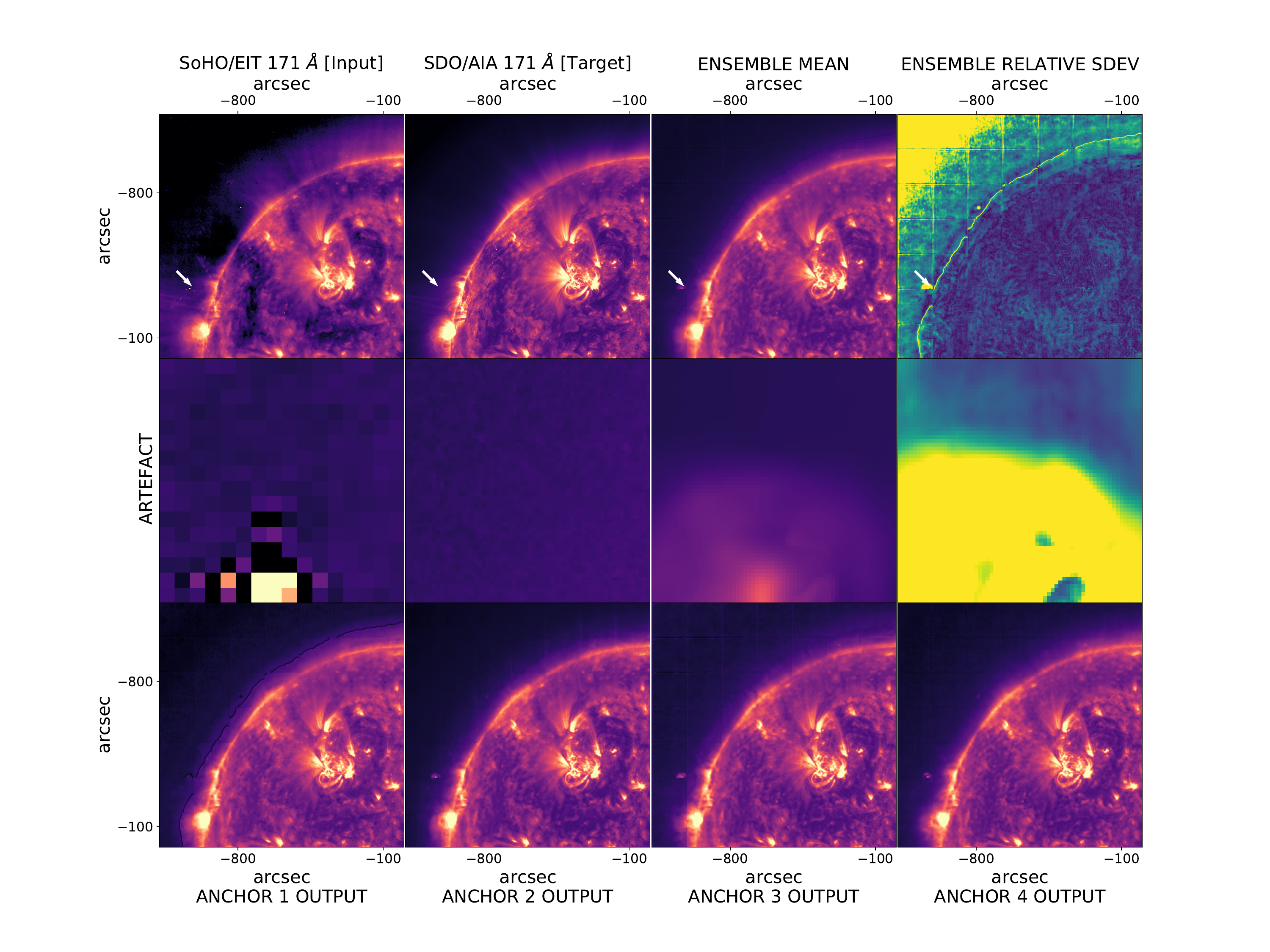}
\caption{Ensemble inference on a single image quadrant of a representative EUV 171~\AA image. The top row starting from left depicts the model input, target, ensemble inference mean, and ensemble relative standard deviation ($\sigma/\mu$). The bottom row shows the outcomes of different ensemble members. It can be observed those bright point-like artifacts (marked by white arrows in the top row and zoomed in the middle row of images) on the input images show high uncertainty and the across ensemble members. Also, the image for relative ensemble standard deviation depicts high uncertainty for the coronal region in the top left corner.}
\label{fig:uncert_3}
\end{figure}



\subsection{Sensitivity of uncertainty to observational artifacts}
On a representative solar image quadrant, we find that regions that are not well represented in the training data (e.g. coronal regions) produce high disagreement among the ensembles. Also, structures of non-solar origin generate high uncertainty (Figure~\ref{fig:uncert_3}).

It is unknown at this moment how many ensemble members are optimal beyond which the uncertainty in reconstruction converges. Further investigations are needed for that.

\section{DISCUSSION AND CONCLUSION}
We thus trained an ensemble of convolutional neural networks to super-resolve SoHO/EIT 171~\AA~images to SDO/AIA 171~\AA~images using the overlapping time period. The salient features of this work are listed below:
\begin{enumerate}
\item We prepared ML-ready data consisting of aligned EIT 171~\AA~$64\times64$~patches and AIA 171~\AA~$256\times256$~patches as input and output images.
\item We trained a CNN based on residual blocks and upsampling layers to translate an EIT patch to an AIA patch
\item We found DL outcome is always superior to a simple upsampling based on intensity scaling and bi-cubic interpolation.
\item We added different terms in the loss function and found having SSIM or Image Gradient as additional terms helps improve the quality of the super-resolved images both visually and quantitatively as compared to MSE baseline.
\item We also estimated the uncertainty of the prediction by training an ensemble of models through Approximate Bayesian Ensembling (ABE).
\item We found that uncertainty improves while making the size of the training set larger.
\item We also found that the model ensemble generated high uncertainty for regions that are not well represented in the training set such as coronal regions, and regions of non-solar origin such as artifacts.
\end{enumerate}

We would like to highlight that ABE has been demonstrated to behave like a Bayesian Neural Network for toy applications for which there is ground truth.  However, there is no Bayesian Neural Network ground truth for homogenization and super-resolution.  Our results are encouraging in that ABE uncertainty has desirable and verifiable properties such as sensitivity to input data, robustness to changes in ensembling parameters, and sensible behavior to unexpected features.

As a future work, we would like to investigate the point of convergence of ensemble uncertainty. Also, translation between unpaired image domains \citep{8237506, 2022mlph.conf...39J} can help make use of the large database beyond overlapping time-period and better constrain high-resolution textural information. We also plan to use our homogenized EUV data for scientific investigations in heliophysics toward independent verification of the data quality (or its scientific readiness) indicated by the metrics.

\section{ACKNOWLEDGEMENTS}
This work is mainly supported by SwRI internal research grant R6134 and NASA HGIO grant 80NSSC23K0416. The authors also acknowledge partial support by NASA SWO2R grant 80NSSC20K0290.


\begin{thebibliography}{13}
\expandafter\ifx\csname natexlab\endcsname\relax\def\natexlab#1{#1}\fi

\bibitem[{Brunelli(2009)}]{10.5555/1643435}
Brunelli, R. 2009, Template Matching Techniques in Computer Vision: Theory and
  Practice (Wiley Publishing)

\bibitem[{{Chatterjee} {et~al.}(2022){Chatterjee}, {Mu{\~n}oz-Jaramillo}, \&
  {Lamb}}]{2022NatAs...6..796C}
{Chatterjee}, S., {Mu{\~n}oz-Jaramillo}, A., \& {Lamb}, D.~A. 2022, Nature
  Astronomy, 6, 796

\bibitem[{{Delaboudini{\`e}re} {et~al.}(1995){Delaboudini{\`e}re}, {Artzner},
  {Brunaud}, {Gabriel}, {Hochedez}, {Millier}, {Song}, {Au}, {Dere}, {Howard},
  {Kreplin}, {Michels}, {Moses}, {Defise}, {Jamar}, {Rochus}, {Chauvineau},
  {Marioge}, {Catura}, {Lemen}, {Shing}, {Stern}, {Gurman}, {Neupert},
  {Maucherat}, {Clette}, {Cugnon}, \& {Van Dessel}}]{1995SoPh..162..291D}
{Delaboudini{\`e}re}, J.~P., {et~al.} 1995, \solphys, 162, 291

\bibitem[{Deudon {et~al.}(2020)Deudon, Kalaitzis, Goytom, Arefin, Lin,
  Sankaran, Michalski, Kahou, Cornebise, \& Bengio}]{deudon2020highresnet}
Deudon, M., {et~al.} 2020, HighRes-net: Recursive Fusion for Multi-Frame
  Super-Resolution of Satellite Imagery

\bibitem[{{Gal} \& {Ghahramani}(2015)}]{2015arXiv150602142G}
{Gal}, Y., \& {Ghahramani}, Z. 2015, arXiv e-prints, arXiv:1506.02142

\bibitem[{Glorot \& Bengio(2010)}]{pmlr-v9-glorot10a}
Glorot, X., \& Bengio, Y. 2010, in Proceedings of Machine Learning Research,
  Vol.~9, Proceedings of the Thirteenth International Conference on Artificial
  Intelligence and Statistics, ed. Y.~W. Teh \& M.~Titterington (Chia Laguna
  Resort, Sardinia, Italy: PMLR), 249--256

\bibitem[{{Jarolim}(2022)}]{2022mlph.conf...39J}
{Jarolim}, R. 2022, in Proceedings of the 2nd Machine Learning in Heliophysics,
  39

\bibitem[{{Jungbluth} {et~al.}(2019){Jungbluth}, {Gitiaux}, {Maloney},
  {Shneider}, {Wright}, {Kalaitzis}, {Deudon}, {G{\"u}ne{\c{s}} Baydin}, {Gal},
  \& {Mu{\~n}oz-Jaramillo}}]{2019arXiv191101490J}
{Jungbluth}, A., {et~al.} 2019, arXiv e-prints, arXiv:1911.01490

\bibitem[{{Lemen} {et~al.}(2012){Lemen}, {Title}, {Akin}, {Boerner}, {Chou},
  {Drake}, {Duncan}, {Edwards}, {Friedlaender}, {Heyman}, {Hurlburt}, {Katz},
  {Kushner}, {Levay}, {Lindgren}, {Mathur}, {McFeaters}, {Mitchell}, {Rehse},
  {Schrijver}, {Springer}, {Stern}, {Tarbell}, {Wuelser}, {Wolfson}, {Yanari},
  {Bookbinder}, {Cheimets}, {Caldwell}, {Deluca}, {Gates}, {Golub}, {Park},
  {Podgorski}, {Bush}, {Scherrer}, {Gummin}, {Smith}, {Auker}, {Jerram},
  {Pool}, {Soufli}, {Windt}, {Beardsley}, {Clapp}, {Lang}, \&
  {Waltham}}]{2012SoPh..275...17L}
{Lemen}, J.~R., {et~al.} 2012, \solphys, 275, 17

\bibitem[{{Pearce} {et~al.}(2018){Pearce}, {Leibfried}, {Brintrup}, {Zaki}, \&
  {Neely}}]{2018arXiv181005546P}
{Pearce}, T., {Leibfried}, F., {Brintrup}, A., {Zaki}, M., \& {Neely}, A. 2018,
  arXiv e-prints, arXiv:1810.05546

\bibitem[{Wang {et~al.}(2018)Wang, Li, Ouyang, \&
  Wang}]{DBLP:journals/corr/abs-1804-09398}
Wang, Z., Li, H., Ouyang, W., \& Wang, X. 2018, CoRR, abs/1804.09398

\bibitem[{{Zhou Wang} {et~al.}(2004){Zhou Wang}, {Bovik}, {Sheikh}, \&
  {Simoncelli}}]{1284395}
{Zhou Wang}, {Bovik}, A.~C., {Sheikh}, H.~R., \& {Simoncelli}, E.~P. 2004, IEEE
  Transactions on Image Processing, 13, 600

\bibitem[{Zhu {et~al.}(2017)Zhu, Park, Isola, \& Efros}]{8237506}
Zhu, J.-Y., Park, T., Isola, P., \& Efros, A.~A. 2017, in 2017 IEEE
  International Conference on Computer Vision (ICCV), 2242--2251
\end{thebibliography}
\end{document}